\documentclass[aps,pra,twocolumn,nofootinbib,a4paper,floatfix,10pt]{revtex4-1}

\usepackage{color}
\usepackage{bm}
\usepackage{amssymb,amsmath,color,graphicx}
\usepackage{amsfonts}
\usepackage{dsfont}
\usepackage{bbold}
\usepackage{graphicx}
\usepackage{float}
\usepackage{subfigure}
\usepackage{verbatim}
\usepackage{amsfonts}
\usepackage{dcolumn}
\usepackage{bm}
\usepackage{color}
\usepackage[dvipsnames]{xcolor}
\usepackage{listings}
\definecolor{zaffre}{rgb}{0.0, 0.08, 0.66}
\usepackage[colorlinks=true,urlcolor=zaffre,citecolor=zaffre,linkcolor=zaffre]{hyperref}
\usepackage{mathrsfs}
\usepackage{filecontents}

\newcommand{\bra}[1]{\mbox{$\langle #1 |$}}
\newcommand{\ket}[1]{\mbox{$| #1 \rangle$}}

\newcommand{\tr}{\mbox{tr}}

\makeatletter
\def\maketag@@@#1{\hbox{\m@th\normalfont\normalsize#1}}  
\makeatother

\begin{document}

\title{Simulation of Gaussian channels via teleportation\\ and error correction of Gaussian states}
\author{Spyros Tserkis}  \email{s.tserkis@uq.edu.au}
\affiliation{Centre for Quantum Computation and Communication Technology, School of Mathematics and Physics, University of Queensland, St Lucia, Queensland 4072, Australia}
\author{Josephine Dias}  \email{josephine.dias@uqconnect.edu.au}
\affiliation{Centre for Quantum Computation and Communication Technology, School of Mathematics and Physics, University of Queensland, St Lucia, Queensland 4072, Australia}
\author{Timothy C. Ralph}  \email{ralph@physics.uq.edu.au}
\affiliation{Centre for Quantum Computation and Communication Technology, School of Mathematics and Physics, University of Queensland, St Lucia, Queensland 4072, Australia}
\date{\today}

\begin{abstract}
Gaussian channels are the typical way to model the decoherence introduced by the environment in continuous-variable quantum states. It is known that those channels can be simulated by a teleportation protocol using as a resource state either a maximally entangled state passing through the same channel, i.e., the Choi-state, or a state that is entangled at least as much as the Choi-state. Since the construction of the Choi-state requires infinite mean energy and entanglement, i.e. it is unphysical, we derive instead every physical state able to simulate a given channel through teleportation with finite resources, and we further find the optimal ones, i.e., the resource states that require the minimum energy and entanglement. We show that the optimal resource states are pure and equally entangled to the Choi-state as measured by the entanglement of formation. We also show that the same amount of entanglement is enough to simulate an equally decohering channel, while even more entanglement can simulate less decohering channels. We, finally, use that fact to generalize a previously known error correction protocol by making it able to correct noise coming not only from pure loss but from thermal loss channels as well.
\end{abstract}

\maketitle

\section{Introduction}

Quantum decoherence is an inevitable feature of any realistic quantum system, leading to errors in the information encoded on it. This decoherence process is mathematically modeled through quantum channels, which induce a corresponding transformation on the states passing through them. Understanding and correcting these errors is the main theoretical and technological barrier to quantum computation and communication overtaking their classical counterparts.

Gaussian states constitute the optimal states for several quantum protocols and are widely used in the photonics community due to the well-established techniques for realizing them, e.g., via optical parametric amplification \cite{Serafini.B.17,Weedbrook.et.al.RVP.12,Adesso.Ragy.OSID.14}. The most typical kind of decoherence induced on those states along their propagation through optical fibers or free space is also Gaussian, and thus being able to error correct this type of noise is vital for quantum communication purposes.

A key tool in quantum information theory is teleportation, originally developed for discrete variables (DV)\cite{Bennett.et.al.PRL.93} and then extended to continuous-variable (CV) systems \cite{Vaidman.PRA.94,Braunstein.Kimble.PRL.98,Ralph.OL.99}. Using entanglement as a resource, teleportation allows a quantum state to be moved from one place to another using classical communication. Realistically, though, teleportation will not lead to a perfect reconstruction, and thus the whole process can be thought of as a quantum channel that induces noise on the initial state \cite{Ralph.Lam.Polkinghorne.JOB.99,Bowen.Bose.PRL.01}. For Gaussian channels, it has been shown that the inverse is also true. Every Gaussian channel can be simulated by a teleportation protocol iff the resource state is either a maximally entangled state passing through the channel that we want to simulate, i.e., the Choi-state, or a state that is at least equally entangled to the Choi-state \cite{Giedke.Cirac.PRA.02,Niset.Fiurasek.Cerf.PRL.09,Holevo.JMP.11,Pirandola.et.al.NC.17,Pirandola.et.al.QST.18}. If, by distillation techniques, a resource state can be established across the channel with more entanglement than the Choi-state, then the simulated teleportation channel may be less decohering than the physical channel, i.e., the state passing through the channel may be error corrected. Quantitatively identifying the resources required for error correction is a key open problem in CV quantum information.

In CV quantum information the Choi-state is not a physical one, since its creation would require infinite energy. Recently it was asked if physical states (states with finite mean energy), with the same entanglement as the Choi-state, can be found that are able to perform channel simulation via teleportation. Logarithmic negativity (see Refs~\cite{Zyczkowski.et.al.PRA.98,Vidal.Werner.PRA.02,Plenio.PRL.05,Adesso.Serafini.Illuminati.PRA.04}) was used as the entanglement quantifier in that analysis \cite{Scorpo.et.al.PRL.17}, and the result was that such states cannot be found for all phase-insensitive Gaussian channels. In particular, pure loss and pure amplifier channels were excluded. 

In this work, we show that this negative result was an artifact of the use of logarithmic negativity as the quantifier of entanglement. Here, using entanglement of formation as the quantifier \cite{Bennett.et.al.PRA.96,Marian.Marian.PRL.08,Akbari-Kourbolagh.Alijanzadeh-Boura.QIP.15,Wolf.et.al.PRA.04,Ivan.Simon.arXiv.08,Tserkis.Ralph.PRA.17}, we prove that all phase-insensitive Gaussian channels can be simulated via teleportation with a physical resource state equally entangled to the Choi-state, and we find the ones with the minimum mean energy. The only exception is the identity, which is expected, since the identity channel represents an ideal teleportation, which by definition requires maximal entanglement, and thus infinite energy. 

We also propose an experimentally accessible way to construct such finite mean energy resource states that can either simulate the initial channel (for pure loss/amplifier channels) or simulate another channel that decoheres the entanglement of formation of an incoming state by the same amount (for thermal loss/amplifier channels). Further, we show that resource states with entanglement more than the Choi-state are able to simulate channels (thermal loss channels as an example) that decohere an incoming state less than the initial channel, and thus error correct the quantum states. This error correction protocol generalizes a previous one which was restricted to pure loss channels \cite{Ralph.PRA.11,Dias.Ralph.PRA.18}.

In sections \ref{secGaussianStates} and \ref{secGaussianChannels} we review Gaussian states and Gaussian channels, respectively, and introduce our nomenclature. Quantification of entanglement is discussed in section \ref{secQuantifyingEntanglement}, where entanglement of formation is defined, and useful expressions for its quantification are derived for relevant Gaussian systems. In section \ref{secQuantumTeleportationandChannelSimulation}, we discuss the connection between quantum teleportation and channel simulation, and we derive analytical expressions for every possible resource state able to simulate a phase-insensitive Gaussian channel. Further, we identify the optimal states, i.e., the ones that need the minimum resources to be constructed, measured in energy and entanglement. Finally, in section \ref{secErrorCorrection}, we introduce and analyze our error correction protocol able to correct Gaussian noise induced on Gaussian states.

\section{Gaussian States}
\label{secGaussianStates}

Let us start by introducing the states that we are going to work with \cite{Serafini.B.17,Weedbrook.et.al.RVP.12,Adesso.Ragy.OSID.14}. A quantum n-mode bosonic state can be described by a vector of the quadrature field operators $\hat{q}:=(\hat{x}_1,\hat{p}_1,\ldots, \hat{x}_n,\hat{p}_n)^T$, with $\hat{x}_j := \hat{a}_j + \hat{a}^{\dag}_j$ and $\hat{p}_j := i(\hat{a}^{\dag}_j - \hat{a}_j)$, where $\hat{a}_j$ and $\hat{a}^{\dag}_j$ are the annihilation and creation operators, respectively, with $[\hat{a}_i{,}{\hat{a}^{\dag}_j}]{=}\delta_{ij}$. 

A quantum state, where the first two moments of $\hat{q}$, i.e., the mean value and the variance, are sufficient for a complete characterization is called Gaussian. In particular, a null mean value (for simplicity) two-mode Gaussian state can be fully described by a real and positive-definite matrix called the covariance matrix, i.e., $\sigma_{ij}=\frac{1}{2}\langle \{ \hat{q}_i, \hat{q}_j\} \rangle$, where $\{,\}$ is the  anticommutator.  In the standard form $\boldsymbol{\sigma}$ is given by \cite{Duan.et.al.PRL.00,Simon.PRL.00}
\begin{equation}
\boldsymbol{\sigma}^{\text{sf}}=\begin{bmatrix}
a & 0 & c_1 & 0 \\
0 & a & 0 & c_2 \\
c_1 & 0 & b & 0 \\
0 & c_2 & 0 & b
\end{bmatrix} \,,
\label{sf}
\end{equation}

Entanglement in CV optical systems is manifested by the correlations of the field operators $\hat{x}$ and $\hat{p}$, and it is typically created by pumping a nonlinear crystal in a non-degenerate optical parametric amplifier. This process is described by a Gaussian unitary known as the two-mode squeezing operator defined as $\hat{S}_2(r) :=\exp[r(\hat{a}\hat{b}- \hat{a}^{\dag}\hat{b}^{\dag})/2]$, where $r \in \mathbb{R}$ is the squeezing parameter (experimentally is often measured in deciBels, i.e., $10 \log_{10}[e^{2r}] \,\text{dB}$). By applying $S_2(r)$ to a couple of vacua, we obtain a pure state called the two-mode squeezed vacuum, with a covariance matrix elements given by $a=b=\frac{1+\chi^2}{1-\chi^2}$ and $c=\frac{2\chi}{1-\chi^2}$, where $\chi=\tanh r \in [0,1)$. Using symplectic transformations, $S$, any covariance matrix can be transformed into 
\begin{equation}
\boldsymbol{\nu}=S \boldsymbol{\sigma} S^T=\nu_{-}\mathds{1} \oplus \nu_{+} \mathds{1} \,,
\end{equation}
where $1\leq \nu_{-} \leq \nu_{+}$ are called symplectic eigenvalues \cite{Vidal.Werner.PRA.02,Serafini.Illuminati.DeSiena.JPB.04}.

\section{Gaussian Channels}
\label{secGaussianChannels}

\begin{figure}[t]
\centering
  \includegraphics[width=\columnwidth]{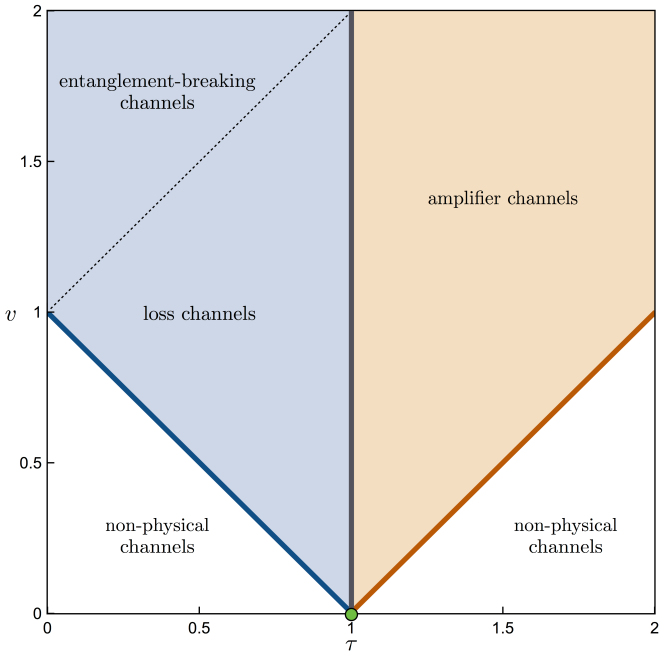}
  \caption{ \small Gaussian Channels. The different classes of phase-insensitive Gaussian channels are presented in this graph. With blue we have the loss channels, $\mathcal{L}$, and the dark blue line indicates the specific case of pure loss channels. With brown we have the amplifier channels, $\mathcal{A}$, and, respectively, the dark brown line represents the pure amplifier channels. The central vertical grey line corresponds to the classical additive noise channels, $\mathcal{N}$, and the green dot indicates the identity channel, $\mathcal{I}$. Channels above the dashed line are entanglement-breaking channels, i.e., $v\geq 1 + |\tau|$, and channels below the dark blue and brown lines are non-physical. All quantities plotted are dimensionless.}
  \label{fig1}
\end{figure}

The decoherence introduced by the environment to a quantum state can be described by a completely positive trace-preserving map called a quantum channel \cite{Serafini.B.17,Weedbrook.et.al.RVP.12,Holevo.PIT.07}. The covariance matrix transformation that a phase-insensitive single-mode Gaussian channel, $\mathcal{G}$, induces in a two-mode Gaussian state is
\begin{equation}
\boldsymbol{\sigma}_{\text{out}} = \mathcal{G}(\boldsymbol{\sigma}_{\text{in}})=(\mathds{1} \oplus \mathcal{U})\boldsymbol{\sigma}_{\text{in}} (\mathds{1}\oplus \mathcal{U})^T + (\mathbb{0} \oplus \mathcal{V})\,,
\end{equation}
where $\mathcal{U}= \sqrt{\tau} \mathds{1}$ and $\mathcal{V}= v \mathds{1}$. Significant phase-insensitive Gaussian channels (see also Fig.~\ref{fig1}) are the following: 
\begin{itemize}
\item loss channel, $\mathcal{L}$, with transmissivity $0 < \tau < 1$ and noise $v=(1-\tau)\varepsilon$ (pure loss for $\varepsilon = 1$, thermal loss for $\varepsilon > 1$),
\item amplifier channel, $\mathcal{A}$, with gain $\tau > 1$ and noise $v=(\tau-1)\varepsilon$ (pure amplifier for $\varepsilon = 1$, thermal amplifier for $\varepsilon > 1$),
\item classical additive noise channel, $\mathcal{N}$, with $\tau = 1$ and noise $v > 0$,
\item identity channel, $\mathcal{I}$, with $\tau=1$ and $v=0$, representing the ideal non-decohering channel.
\end{itemize}

A loss channel, $\mathcal{L}$, is one which can be modeled as a beam-splitter operation, $\hat{B}:=\exp[\varphi(\hat{a}^{\dag}\hat{b}- \hat{a}\hat{b}^{\dag})]$, with transmissivity $\tau=\cos^2 \varphi$, and a vacuum (shot noise given by $\varepsilon=1$) or a thermal state ($\varepsilon > 1$) at the other input for pure or thermal loss, respectively. Similarly, a pure or thermal amplifier channel, $\mathcal{A}$, is modeled by a two-mode squeezing operation (defined in section \ref{secGaussianStates}) with gain $\tau=\cosh 2r$ and a vacuum or a thermal state at the other input, respectively \cite{Caves.PRD.82}. Classical additive noise channel, $\mathcal{N}$, is an asymptotic case of either loss or thermal channels where $\tau \approx 1$ and a highly thermal state, i.e., classical, at the other input. Finally, the identity channel, $\mathcal{I}$, is an ideal case where the transmissivity/ gain is unity, $\tau = 1$, so there is no interaction with the environment and no noise is induced.

\section{Quantifying Entanglement}
\label{secQuantifyingEntanglement}

Entanglement in pure states is measured by the entropy of entanglement defined as $\mathcal{E}(\ket{\psi}) := \mathcal{S} (\tr_B \ket{\psi}\bra{\psi})$, where $\mathcal{S}(x):=-\tr (x \log_2 x)$ is the von Neumann entropy, and $\tr_B$ denotes the partial trace over subsystem $B$. For mixed states, there is a plethora of measures, which, in general do not coincide with each other. A proper entanglement measure for mixed states is entanglement of formation, defined as the convex-roof extension of the von Neumann entropy, $\mathcal{E} :=\inf \{\sum_i p_i \mathcal{S} (\tr_B \ket{\psi_i} \bra{\psi_i})\}$ \cite{Bennett.et.al.PRA.96}.

For two-mode Gaussian states \cite{Marian.Marian.PRL.08,Akbari-Kourbolagh.Alijanzadeh-Boura.QIP.15,Wolf.et.al.PRA.04,Ivan.Simon.arXiv.08,Tserkis.Ralph.PRA.17} entanglement of formation is given by
\begin{equation}
\mathcal{E} (\boldsymbol{\sigma}) := \cosh^2 r_o \log_2 ( \cosh^2 r_o)-\sinh^2 r_o \log_2 ( \sinh^2 r_o)\,,
\end{equation}
with $r_o$ representing the minimum two-mode squeezing required to prepare a state $\boldsymbol{\sigma}$ \cite{Tserkis.Ralph.PRA.17}. When a single arm of a two-mode squeezed vacuum goes through a channel $\mathcal{G}$ with parameters $\tau$ and $v$, $r_o$ is given by
\begin{align}
r_o{=}&\frac{1}{4}\ln \frac{1}{2 \left[v{-}2 \sqrt{\tau } \sinh (2 r){+}(1{+}\tau) \cosh (2 r)\right]^2}  \nonumber \\
& \hspace{0.6cm} {\times} \Big\{ 3{+} \left[2 v^2{-}(1{-}\tau )^2 \right] \cosh(4 r)  \nonumber \\
& \hspace{1.1cm} {+} \tau (3 \tau {+}2){+} 4 v (1{+}\tau) \cosh (2 r)  \nonumber \\
& \hspace{1.1cm} {-} 4 \sqrt{v^2{-}(1{-}\tau )^2} \sinh (2 r) [v \cosh (2 r){+}1{+}\tau] \Big\} ,
\end{align}
while for the corresponding case of an infinitely squeezed state ($\chi=\tanh r \rightarrow 1$), i.e., the Choi-state \cite{Giedke.Cirac.PRA.02,Holevo.JMP.11}, we have
\begin{subequations}
\begin{align}
\tau& \neq 1 \; \Rightarrow \; r_o^{\text{Choi}}=\frac{1}{4}\ln \frac{2 v \left[v{-}\sqrt{v^2{-}(1{-}\tau )^2}\right]{-}(1{-}\tau)^2}{\left(1{-}\sqrt{\tau }\right)^4}, \label{choikappa1} \\
\tau& = 1 \; \Rightarrow \;  r_o^{\text{Choi}}=\frac{1}{4}\ln \frac{4}{v^2}\,. \label{choikappa2}
\end{align}
\end{subequations}

Note that for $v\geq 1 + |\tau|$, we have an entanglement-breaking channel \cite{Namiki.Hirano.PRL.04,Holevo.PIT.08}, i.e., the entanglement vanishes and $r_o=0$ by definition.

\section{Quantum Teleportation and Channel Simulation}
\label{secQuantumTeleportationandChannelSimulation}

Let us assume that we want to teleport a single-mode of a (null mean valued) two-mode Gaussian state with covariance matrix $\boldsymbol{\sigma}_{\text{in}}$ from one place (Lab 1) to another (Lab 2). We need an entangled state shared between Lab 1 and Lab 2, called the resource state, i.e., a state with a covariance matrix given by
\begin{equation}
\boldsymbol{\rho}=\begin{bmatrix}
a & 0 & c & 0 \\
0 & a & 0 & -c \\
c & 0 & b & 0 \\
0 & -c & 0 & b
\end{bmatrix} \,.
\label{resource}
\end{equation}

\begin{figure}[t]
\centering
  \includegraphics[width=\columnwidth]{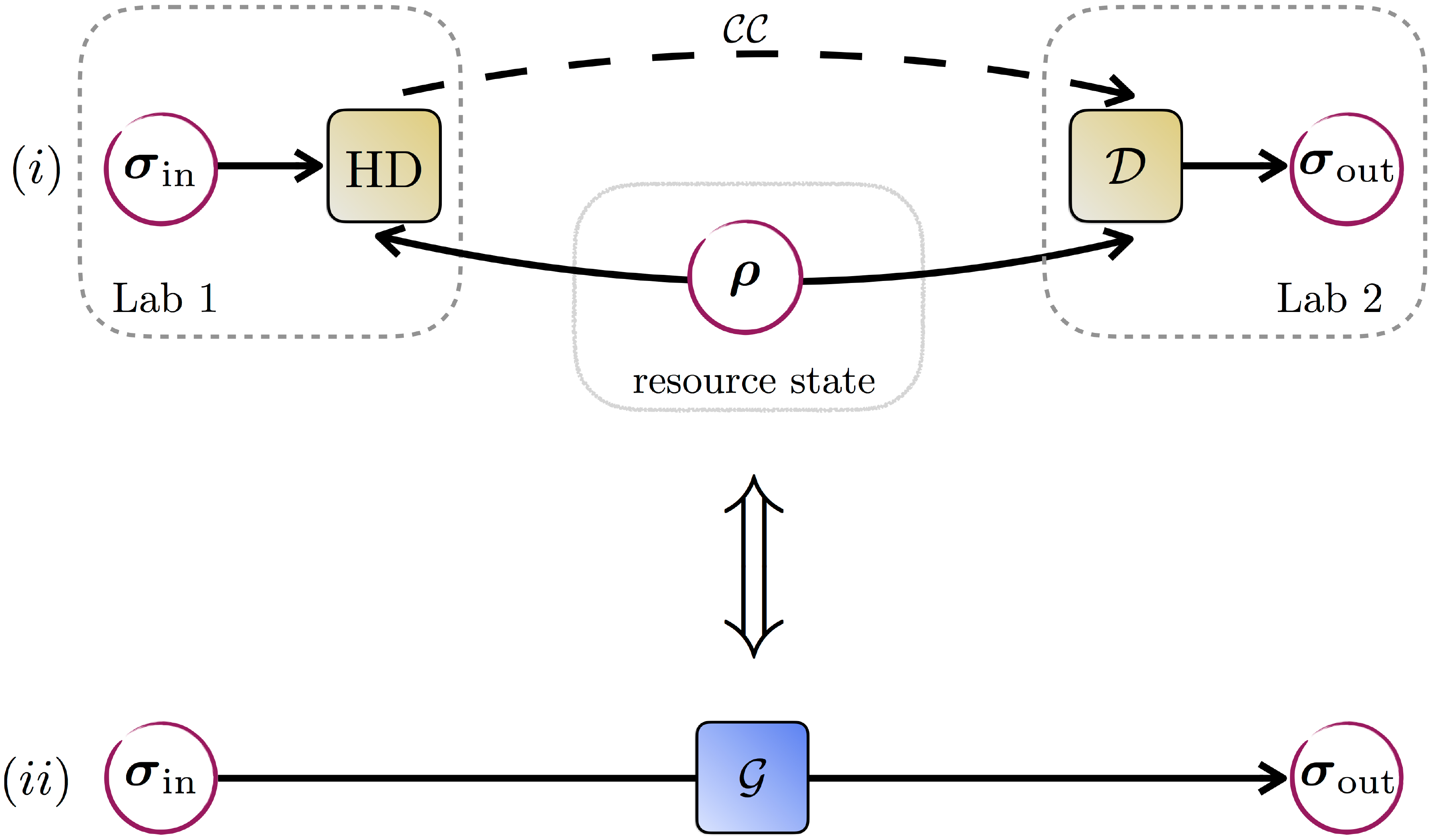}
  \caption{ \small Teleportation and channel simulation. The teleportation protocol \cite{Braunstein.Kimble.PRL.98} is represented in figure $(i)$ via its basic components: a) the dual homodyne detection, HD, between the resource state, $\boldsymbol{\rho}$, and the initial state $\boldsymbol{\sigma_{\text{in}}}$, b) the classical channel, $\mathcal{CC}$, c) the displacement, $\mathcal{D}$, and d) the output state, $\boldsymbol{\sigma_{\text{out}}}$. In figure $(ii)$ we have the corresponding channel that the teleportation protocol simulates \cite{Ralph.Lam.Polkinghorne.JOB.99}.}
  \label{fig2}
\end{figure}

In Lab 1, one arm of the resource state is mixed with the input state through a balanced beam-splitter, followed by a dual homodyne detection (measuring $\hat{x}$ on one arm and $\hat{p}$ on the other), HD, and the results of that measurement are sent to Lab 2 through a classical channel, $\mathcal{CC}$. Finally, in Lab 2, a displacement operation proportional to the results of these measurements, $\mathcal{D}$, is applied on the other arm of the resource state in order to reconstruct the input state, i.e., teleport it, $\boldsymbol{\sigma}_{\text{out}}$. The above protocol, presented graphically in Fig.~\ref{fig2}(i), has been introduced in Ref.~\cite{Braunstein.Kimble.PRL.98}, while a review of CV teleportation can be found in Ref.~\cite{Pirandola.Mancini.LP.06}. 

From a mathematical point of view, teleportation is equivalent to a Gaussian channel transformation \cite{Ralph.Lam.Polkinghorne.JOB.99}, as pictured in Fig.~\ref{fig2}(ii). Using the balanced correlated resource state given in Eq.~\ref{resource}, we get a phase-insensitive channel with:
\begin{equation}
\tau=\lambda\,, \quad \quad \quad v= a \lambda - 2 c \sqrt{\lambda}+b \,,
\label{chsim}
\end{equation}
where $\lambda \geq 0$ is the experimentally accessible gain of the classical channel. Conversely, Gaussian channels can always be simulated via teleportation by using as a resource state either the Choi-state or a finite mean energy state, $\boldsymbol{\rho}$, which is at least equally entangled to the Choi-state \cite{Giedke.Cirac.PRA.02,Niset.Fiurasek.Cerf.PRL.09,Pirandola.et.al.NC.17,Pirandola.et.al.QST.18}
\begin{equation}
\mathcal{E}(\boldsymbol{\rho}) \geq \mathcal{E}(\mathcal{G}^{\text{Choi}})\,.
\label{choieq}
\end{equation}

Since the Choi-state is an unphysical one, our goal is to find a physical resource state, $\boldsymbol{\rho}$, able to simulate a Gaussian channel $\mathcal{G}$ through teleportation. In order to do so, we assume a resource state of the form given in Eq.~\ref{resource} that satisfies Eq.~\ref{chsim}. It is also convenient to express the symplectic eigenvalues through the covariance matrix elements, i.e., $\nu_{\pm}=\frac{\sqrt{(a+b)^2-4c^2} \pm |a-b| }{2}$ \cite{Serafini.Illuminati.DeSiena.JPB.04}. Then, solving this system for $a$ , $b$ and $c$, and assuming $a \geq b$, we end up with a resource state with the following covariance matrix elements:
\begin{widetext}
\begin{subequations}
\begin{align}
a_{\pm}&=\frac{ (1-\tau) (\nu_{+}-\nu_{-}) + (1 + \tau)v \pm 2 \sqrt{\tau  (\tau\nu_{-} - \nu_{-}+v) (\nu_{+}-\tau \nu_{+} +v)}}{(\tau - 1)^2} \,, \label{physicalstates1} \\
b_{\pm}& = \frac{ \tau (1-\tau) (\nu_{+}-\nu_{-})  +  (1 + \tau)v \pm 2 \sqrt{\tau  (\tau\nu_{-} - \nu_{-} + v) (\nu_{+} - \tau \nu_{+} +v)}}{(\tau - 1)^2} \,, \label{physicalstates2}\\
c_{\pm}& = \frac{  \tau (1-\tau) (\nu_{+}-\nu_{-})  + 2 \tau  v  \pm (1 + \tau) \sqrt{\tau  (\tau\nu_{-} - \nu_{-} + v) (\nu_{+}-\tau \nu_{+} +v)} }{\sqrt{\tau }(\tau - 1)^2 } \,, \label{physicalstates3}
\end{align}
\end{subequations}
\end{widetext}
which give two set of states, i.e., $\boldsymbol{\rho}_{\pm}$. For finite values of $\nu_{\pm}$ we get every physical state that can simulate a given channel $\mathcal{G}$ with parameters $\tau$ and $v$. However, for $\nu_{-}(\boldsymbol{\rho}_{-})=\nu_{+}(\boldsymbol{\rho}_{-})=1$, both entanglement, $\mathcal{E}(\boldsymbol{\rho}_{\pm})$, and mean energy per mode, $\langle \hat{a}^{\dag} \hat{a} \rangle_{\boldsymbol{\rho}_{\pm}}=\frac{\tr(\boldsymbol{\rho}_{\pm})-4}{8}$, are minimized, which corresponds to pure states with squeezing parameter equal to
\begin{equation}
\chi_{\text{opt}}=\frac{2 \sqrt{\tau }-\sqrt{(v+1-\tau) ( v-1+\tau)}}{\tau +v+1} \,.
\label{optstates}
\end{equation}

The entanglement of those states is exactly equal to the corresponding Choi-state, $\mathcal{E}(\boldsymbol{\rho}_{\text{opt}}) = \mathcal{E}(\mathcal{G}^{\text{Choi}})$, which is the minimum possible needed for the simulation, saturating Eq.~\ref{choieq}. Thus, they are the optimum resource states for channel simulation.

Note that states of the form of Eq.~\ref{optstates} were considered in Ref.~\cite{Scorpo.et.al.PRL.17} for channel simulation, but since the analysis in that paper was based on logarithmic negativity as the entanglement quantifier, the conclusion was that, even though energetically preferable, they are more entangled than the corresponding Choi-state, according to this specific measure. For that reason, mixed states with higher values of mean energy (see Eq.~9 in Ref.~\cite{Scorpo.et.al.PRL.17} that corresponds to $\nu_-=1$ and $\nu_+ >1$) are considered the optimal ones, since their logarithmic negativity is equal to the desirable one. One further restriction using logarithmic negativity is that pure loss and pure amplifier channels cannot be simulated with resource states equally entangled to the Choi-state (the identity channel is also excluded since it is fundamentally impossible to be simulated with finite energy). 

In our analysis, on the other hand, using entanglement of formation as the quantifier, we find that the optimal resource states (given by Eq.~\ref{optstates}) are equally entangled to the corresponding Choi-state, they have the minimum possible energy (since they are pure states), and they can simulate all phase-insensitive channels (including pure loss and pure amplifier but of course excluding the identity). Let us also mention that, by definition, entanglement of formation quantifies the minimum entanglement needed to create a state, and thus we consider it as the appropriate measure for calculating the minimum required resources in channel simulation as well.

\begin{figure}[t]
\centering
  \includegraphics[width=\columnwidth]{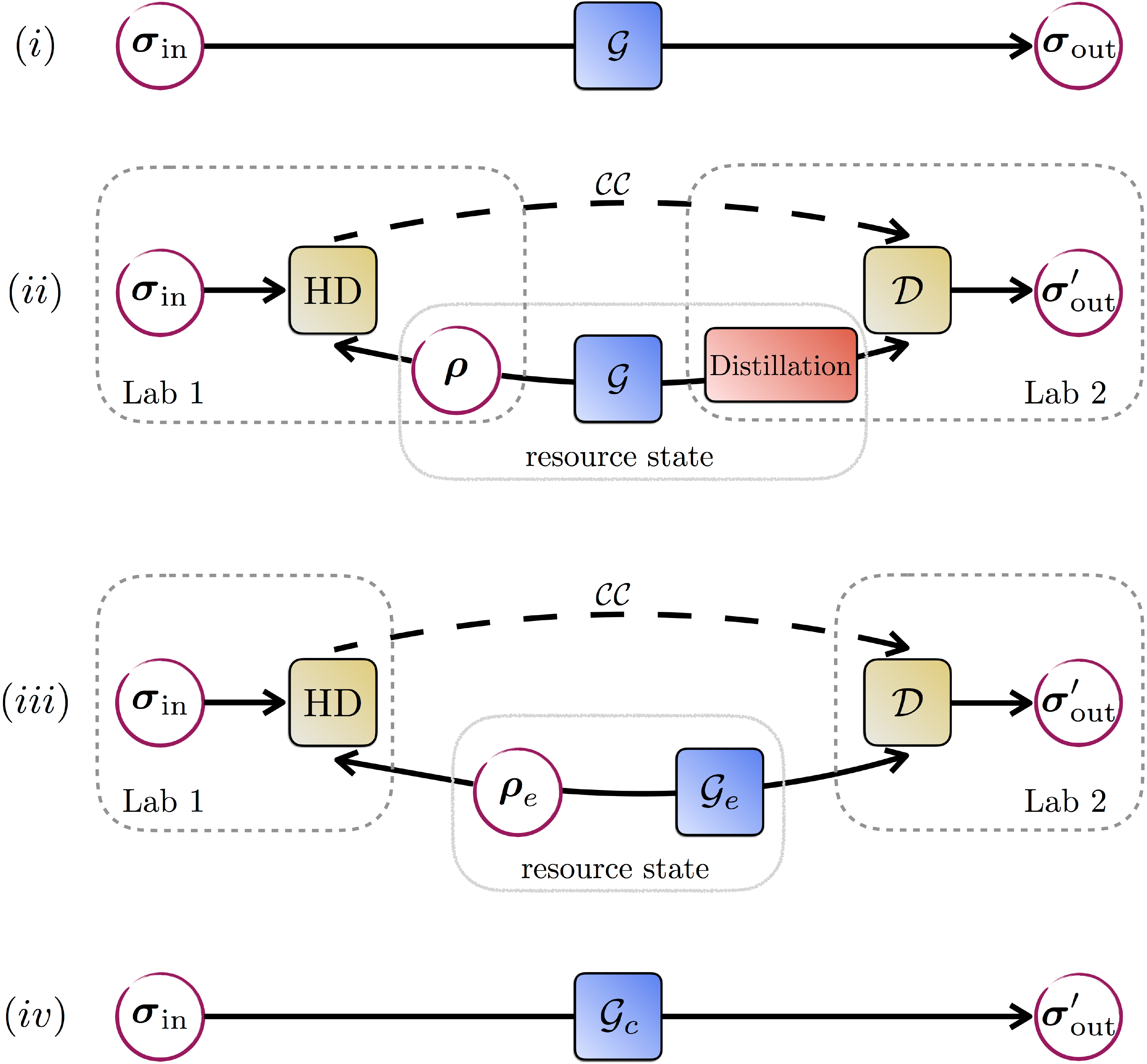}
  \caption{ \small Channel simulation through entanglement distillation. Figure $(i)$ represents the channel, $\mathcal{G}$, that we want to simulate. Figure $(ii)$ shows the resource state $\boldsymbol{\rho}$, which is sent through the same channel, $\mathcal{G}$, and then through an entanglement distillation process before it is used in a teleportation protocol (see also Fig.~\ref{fig2}). In figure $(iii)$ we have the effective transformation for a successful distillation process, i.e., $\mathcal{G}_e$ and $\boldsymbol{\rho}_{e}$, and finally in figure $(iv)$ the simulated channel $\mathcal{G}_c$, which leads to a state $\boldsymbol{\sigma}'_{\text{out}}$.}
  \label{fig3}
\end{figure}

Finding the optimal resource state is a theoretical result that gives us physical insight, but has limited practical application, since, in a realistic scheme, physical limitations exist in both the creation and the transmission of this state. In particular, let us assume that Lab 1 and Lab 2 are separated by a channel $\mathcal{G}$. Instead of sending directly a state through this channel, the two Labs can apply a teleportation protocol. However, the resource state that Lab 1, for instance, has created needs necessarily to pass through the same decohering channel that we started with, $\mathcal{G}$, since this is the environmental decoherence that is beyond our control. This channel, though, will decrease the entanglement of the resource state leading to a simulated channel, in general different from the initial one. For that reason, Lab 2 needs to apply an entanglement distillation protocol in order that the entanglement of the resource state can be enough for the desired simulation. Interestingly, Lab 2 can distill the resource state even more and thus make the teleportation protocol simulate a less decohering channel, $\mathcal{G}_c$, that practically error corrects the state that we wanted to send through channel $\mathcal{G}$. The whole process is schematically illustrated, step by step, in Fig.~\ref{fig3}. In the next section we discuss this realistic construction of the resource state and the error correction protocol in detail.

\section{Error Correction}
\label{secErrorCorrection}

Error correction is a process  based on an added redundancy, induced by either (i) embedding the state into a multipartite entangled state or (ii) teleporting the state using multiple distilled entangled pairs. The equivalence of those two procedures (at least for DV systems) has been shown in \cite{Bennett.et.al.PRA.96} and several protocols have been developed for both DV \cite{Bennett.et.al.PRA.96,Shor.PRA.95,Steane.PRL.96} and CV states \cite{Lloyd.Slotine.PRL.98,Braunstein.N.98,Niset.Andersen.Cerf.PRL.08,Ralph.PRA.11,Gottesman.Kitaev.Preskill.PRA.01,Lund.Ralph.Haselgrove.PRL.08,Ralph.PRA.11,Leghtas.et.al.PRL.13}.

An ideal protocol would totally reconstruct the state by removing all induced errors, so the fidelity \cite{Uhlmann.RMP.76,Marian.Marian.PRA.12,Banchi.Braunstein.Pirandola.PRL.15} between the input and output state (the squared overlap between the corresponding probability distributions \cite{Barnum.et.al.PRL.96}) of the protocol would be equal to 1. In principle, this is feasible in DV codes, since during the distillation process we construct maximally entangled states, i.e., Bell states. In CV codes, focused on Gaussian noise, fidelity equal to 1 is impossible, since the corresponding maximally entangled states, i.e., EPR states, are unphysical. Comparing non-unity fidelities is in general a meaningless task, since the overlap between two probability distributions fails to take into account important features of the state, e.g., classicality, separability, Gaussianity etc. A thorough analysis of this issue can be found in Ref.~\cite{Mandarino.et.al.IJQI.14}, while for the specific case of teleportation in Ref.~\cite{Ralph.Lam.Polkinghorne.JOB.99}. A characteristic example is presented in the Appendix \ref{ap1}.  

Since full reconstruction of the initial state is impossible, our goal is to create a channel that induces less decoherence than the initial one. In order to do so, we apply the scheme illustrated in Fig.~\ref{fig3}. In particular, let us assume that in Lab 1 we have a Gaussian two-mode squeezed vacuum state $\boldsymbol{\sigma_{\text{in}}}$, and we want to send one arm of it to Lab 2 through a channel $\mathcal{G}$ that decoheres it, giving the output state $\boldsymbol{\sigma_{\text{out}}}$. This decoherence leads to a noisy version of the initial state so our task is to remove this noise as much as possible. This error correction process is based on the teleportation protocol described in the previous section \ref{secQuantumTeleportationandChannelSimulation}. 

\paragraph*{Error correction protocol.} Lab 1 creates an entangled state $\boldsymbol{\rho}$, keeping one arm (assumed not decohered) and sending the other to Lab 2 through channel $\mathcal{G}$, which models the unavoidable decoherence due to environmental noise. That channel reduces the entanglement of state $\boldsymbol{\rho}$, so Lab 2 performs a distillation protocol to increase it. Distillation protocols are in general probabilistic, so assuming a successful protocol this step can be modeled as an effective entangled state $\boldsymbol{\rho}_e$ passing through an effective channel $\mathcal{G}_e$. Since Lab 1 and 2 have now established an entangled state $\mathcal{G}_e(\boldsymbol{\rho}_e)$, they use it as a resource state to teleport $\boldsymbol{\sigma_{\text{in}}}$. The more we distill the entanglement of the resource state the less decohering is the simulated channel, so at some point we can create an output state $\boldsymbol{\sigma}'_{\text{out}}$ less decohered (less noisy) than the output state that we would get without the protocol, $\boldsymbol{\sigma_{\text{out}}}$.

\subsection{Pure Channels}

For pure channels an equally decohering channel would necessarily be identical with the initial one, since they are only transmissivity/gain dependent. This can be achieved by solving Eq.~\ref{chsim}, for $v=\pm(1 - \lambda)$ (plus sign for loss and minus for amplification). Assuming a successful distillation we get an effective state $\mathcal{G}_e(\boldsymbol{\rho}_e)$ (see Fig.~\ref{fig3}(ii) and Fig.~\ref{fig3}(iii)), and using this effective state as a resource state, shared between Lab 1 and 2, we apply a teleportation protocol in order to get a simulated channel, $\mathcal{G}_c$ (see Fig.~\ref{fig3}(iv)). The simulated channel due to the effective resource state $\mathcal{G}_e(\boldsymbol{\rho}_e)$, with covariance matrix elements $a_e=\frac{1+\chi_e^2}{1-\chi_e^2}$, $b_e=\tau_e \frac{1+\chi_e^2}{1-\chi_e^2}\pm(1-\tau_e)$ and $c_e=\frac{2\sqrt{\tau_e} \chi_e}{1-\chi_e^2}$, is
\begin{subequations}
\begin{align}
\mathcal{L}_c&\equiv \mathcal{L} \quad \text{with} \quad \tau_c=\lambda=\tau_e \chi_e^2\,,  \\
\mathcal{A}_c&\equiv \mathcal{A} \quad \text{with} \quad \tau_c=\lambda=\tau_e/\chi_e^2\,.
\end{align}
\end{subequations}

A trivial way to achieve that is when the resource state is the Choi-state, i.e., $\chi_e \rightarrow 1$ and $\tau_e =\tau$. However, any resource state with parameters such that $\tau_e \chi_e^2=\tau$ for loss channels and $\tau_e/\chi_e^2=\tau$ for amplifier, would be equally entangled to the Choi-state, i.e., $\mathcal{E}[\mathcal{L}_e(\boldsymbol{\rho}_e)]=\mathcal{E}(\mathcal{L}^{\text{Choi}})$ and $\mathcal{E}[\mathcal{A}_e(\boldsymbol{\rho}_e)]=\mathcal{E}(\mathcal{A}^{\text{Choi}})$, respectively, leading to the same amount of distributed entanglement, $\mathcal{E}(\boldsymbol{\sigma}'_{\text{out}})=\mathcal{E}(\boldsymbol{\sigma}_{\text{out}})$. 

In order to induce less decoherence, i.e., $\mathcal{E}(\boldsymbol{\sigma}'_{\text{out}}) > \mathcal{E}(\boldsymbol{\sigma}_{\text{out}})$, and start the error correction, we would need $\lambda=\tau_c>\tau$, which requires resource states with entanglement greater that the corresponding Choi-state, i.e., $\mathcal{E}[\mathcal{L}_e/\mathcal{A}_e(\boldsymbol{\rho}_e)]>\mathcal{E}(\mathcal{L}^{\text{Choi}}/\mathcal{A}^{\text{Choi}})$.  

\subsection{Thermal Channels}

Assuming now that we have a thermal channel, we have to solve again Eq.~\ref{chsim}, for $v={\pm}(1- \lambda)\theta$ (plus sign for thermal loss and minus for thermal amplification), with $\theta \geq1$, using the corresponding effective covariance matrix elements: $a_e=\frac{1+\chi_e^2}{1-\chi_e^2}$, $b_e=\tau_e \frac{1+\chi_e^2}{1-\chi_e^2}\pm(1-\tau_e)\varepsilon_e$ and $c_e=\frac{2\sqrt{\tau_e} \chi_e}{1-\chi_e^2}$. The teleportation part of the protocol simulates the following quantum channel:
\begin{subequations}
\begin{alignat}{2}
& 0 <  \lambda \leq 1 \;  &\Rightarrow& \quad \mathcal{L}_c\quad \text{with} \quad \{\tau_c = \lambda,\varepsilon_c=\theta\},  \\
&1 \leq \lambda \leq \lambda_{\text{max}} \; &\Rightarrow& \quad \mathcal{A}_c \quad \text{with} \quad \{\tau_c = \lambda,\varepsilon_c=-\theta\} ,
\end{alignat}
\end{subequations}
where
\begin{equation*}
\theta=\frac{\varepsilon_e  (\tau_e {-}1) \left(\chi_e ^2{-}1\right){+}\chi_e ^2 (\tau_e {+}\lambda){-}4 \chi_e \sqrt{\tau_e \lambda } {+}\tau_e {+}\lambda}{(\lambda {-}1) \left(\chi_e ^2{-}1\right)} ,
\end{equation*}
and
\begin{align*}
\lambda_{\text{max}}=&\frac{(\varepsilon_e {+}1) (1{-}\tau_e )}{2}  \nonumber \\
& + \frac{3 \tau_e  {-} \varepsilon_e  (1{-}\tau_e) {-}1{-}2 \sqrt{2 \tau_e(\varepsilon_e {+}1)  (1{-}\tau_e ) \left(\chi_e ^2{-}1\right)}}{2\chi_e ^2} .
\end{align*}

Simulating an identical thermal channel, i.e., $\mathcal{L}_c/\mathcal{A}_c \equiv \mathcal{L}/\mathcal{A}$, can be achieved only in the unphysical limit of the Choi-state (using the above effective resource state). However, if we focus only on the decoherence, we can always create thermal channels that induce equal or less decoherence than the initial one, i.e., $\mathcal{E}(\boldsymbol{\sigma}'_{\text{out}}) \geq \mathcal{E}(\boldsymbol{\sigma}_{\text{out}})$, as long as $\mathcal{E}[\mathcal{L}_e/\mathcal{A}_e(\boldsymbol{\rho}_e)] \geq \mathcal{E}(\mathcal{L}^{\text{Choi}}/\mathcal{A}^{\text{Choi}})$.

\subsection{Error Correction with ideal NLA}

Experimentally, the most challenging part of the protocol is the distillation of entanglement, that induces the effective parameters needed for the above simulations. A distillation technique useful for CV systems is noiseless linear amplification (NLA) \cite{Ralph.Lund.QCMC.09,Xiang.Ralph.Lund.Walk.Pryde.NP.10}. For a loss channel, the effective parameters $\{ \chi_e,\tau_e,\varepsilon_e\}$ have been computed in Refs.~\cite{Walk.Lund.Ralph.13,Blandino.et.al.PRA.15}, but can also be found in the Appendix \ref{ap2}. Before we apply this technique, let us give a brief description on how NLA works.

NLA is a probabilistic procedure, described (in an idealized fashion) by the operator $g^{\hat{n}}$, that implements the number state transformation, 
\begin{equation}
g^{\hat{n}} \ket{n} \rightarrow g^n \ket{n}\,,
\end{equation}
where $g \geq 1$ is an experimentally accessible gain. Interestingly, NLA can be used to distill entanglement, since when, for instance, it is applied to a two-mode squeezed vacuum, it probabilistically gives, 
\begin{equation}
g^{\hat{n}} \sqrt{1-\chi^2} \sum \chi^n \ket{n n}\rightarrow \sqrt{1-\chi^2} \sum g^n \chi^n \ket{n n} \,.
\end{equation}

Thus, when the NLA succeeds, the squeezing parameter is increased, i.e., $\chi \rightarrow g \chi$, which implies entanglement distillation. 

For pure loss channels, using NLA, it has already been shown \cite{Ralph.PRA.11,Dias.Ralph.PRA.18} that in order to simulate equally decohering channels, i.e., $\tau_c=\tau \Rightarrow \mathcal{E}(\boldsymbol{\sigma}'_{\text{out}})=\mathcal{E}(\boldsymbol{\sigma}_{\text{out}})$ we need $g=1/\chi$ and $\lambda=\tau_e \chi_e^2$, while for $g\geq 1/\chi$ we get $\tau_c \geq \tau$ and thus $\mathcal{E}(\boldsymbol{\sigma}'_{\text{out}}) \geq \mathcal{E}(\boldsymbol{\sigma}_{\text{out}})$, so we start the error correction. 

\begin{figure*}[t]
  \centering
  \subfigure{\includegraphics[scale=0.266]{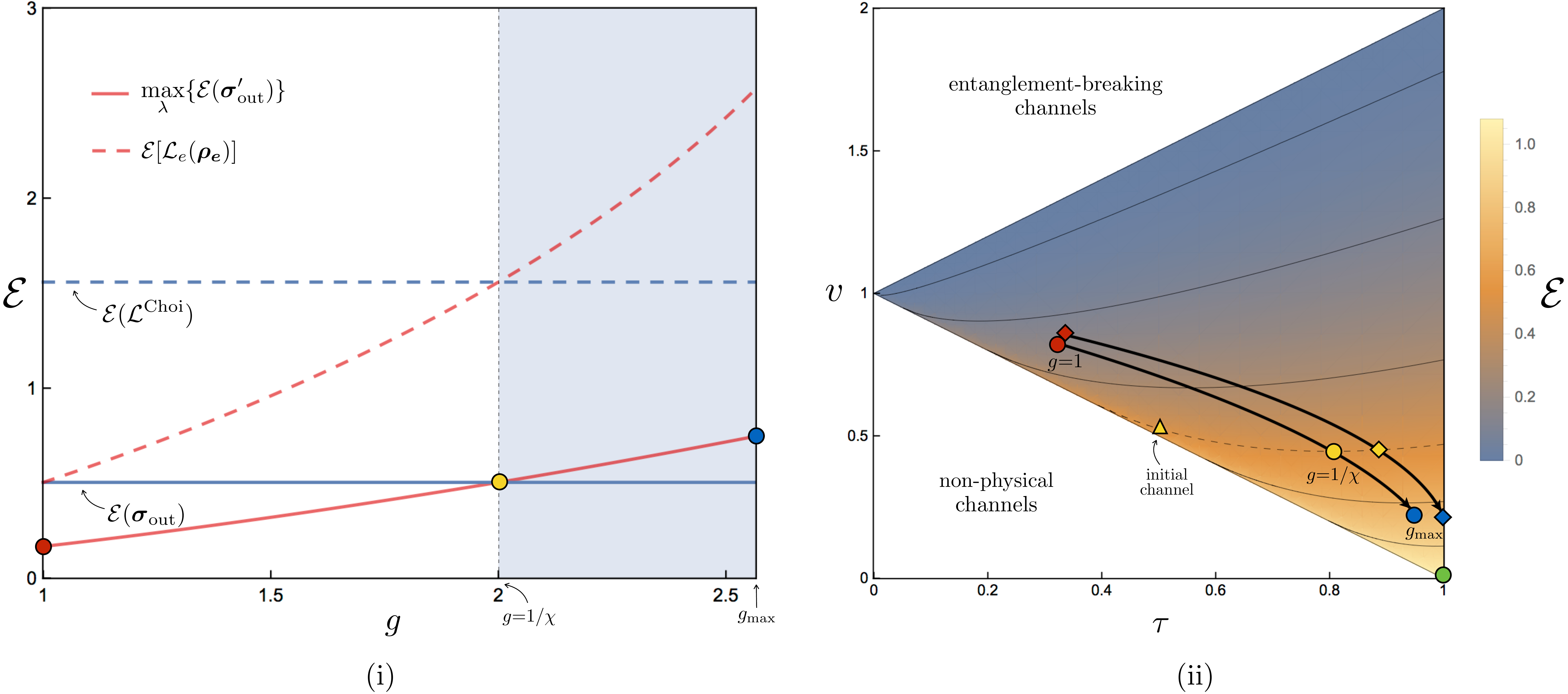}}
  \caption{ \small Error correction with ideal NLA. An initial thermal loss channel with $\tau=0.5$ and $\varepsilon=1.05$ induces noise into one arm of a pure two-mode squeezed state $\boldsymbol{\sigma_{\text{in}}}$ with squeezing parameter $\zeta=0.5$. We apply the protocol using a resource state $\boldsymbol{\rho}$ with squeezing parameter $\chi=0.5$. In figure (i) we present both the (optimized over teleportation gain) entanglement of the output state, $\max_{\lambda}\{\mathcal{E}(\boldsymbol{\sigma}'_{\text{out}})\}$, (red solid line), and the entanglement of the distilled resource state, $\mathcal{E}[\mathcal{L}_e(\boldsymbol{\rho}_e)]$, (red dashed line), against the NLA gain, $g$. With solid blue and dashed blue lines we have the entanglement of the output state without the protocol, $\mathcal{E}(\boldsymbol{\sigma}_{\text{out}})$, and the deterministic upper bound of entanglement for this channel, $\mathcal{E}(\mathcal{L}^{\text{Choi}})$, respectively. We observe that for $g=1/\chi$ the entanglement of the Choi-state is reached and from then on and until we reach $g_{\text{max}}$ we are into the error correction area (light blue shaded), i.e., $\max_{\lambda}\{\mathcal{E}(\boldsymbol{\sigma}'_{\text{out}})\} \geq \mathcal{E}(\boldsymbol{\sigma}_{\text{out}})$. In figure (ii) the contour lines indicate equally decohering channels (with parameters $\tau$ and $v$), i.e., channels that decohere the entanglement by the same amount. The yellow triangle represents the initial channel. Applying the protocol without distilling the resource state, i.e., $g=1$, we get the channel shown with the red dot. Increasing the NLA gain, and specifically for $g=1/\chi$, we simulate a channel (yellow dot in the graph), that decoheres the state by the same amount as the initial channel (both channels lie on the same dashed contour line). The best channel we can simulate is achieved for $g_{\text{max}}$, and is represented by the blue dot. With red/yellow/blue diamonds we indicate the corresponding simulated channels of an error correcting protocol based on a less entangled resource state, i.e., $\chi'=0.45$. Thus, we can visually interpret error correction as the process of simulating a channel ``closer" to the identity (represented by the green dot) than the initial one. All quantities plotted are dimensionless.}
  \label{fig4}
\end{figure*}

For thermal loss channels, in order to reach that bound, we need the same NLA gain, i.e., $g=1/\chi$, but we also have to optimize over the teleportation gain, i.e., $\max_{\lambda}\{\mathcal{E}(\boldsymbol{\sigma}'_{\text{out}})\}=\mathcal{E}(\boldsymbol{\sigma}_{\text{out}})$. As expected, $g=1/\chi$ is also the point when the resource state has entanglement equal to the Choi-state, i.e., $\mathcal{E}[\mathcal{L}_e(\boldsymbol{\rho}_e)]=\mathcal{E}(\mathcal{L}^{\text{Choi}})$. For NLA gain greater than $1/\chi$, we get $\mathcal{E}[\mathcal{L}_e(\boldsymbol{\rho}_e)]>\mathcal{E}(\mathcal{L}^{\text{Choi}})$ and thus $\max_{\lambda}\{\mathcal{E}(\boldsymbol{\sigma}'_{\text{out}})\} \geq \mathcal{E}(\boldsymbol{\sigma}_{\text{out}})$, which implies error correction. A specific example is presented in Fig.~\ref{fig4}.

NLA gain is upper bounded by a finite value $1\leq g \leq g_{\text{max}}$, beyond which the output state becomes unphysical. There are two conditions here that we want to satisfy in order the output distilled state to be physical. The first one is $0 \leq \chi_{e} < 1$, which corresponds to 
\begin{equation} 
1 \leq g \leq  g_{\chi}=\sqrt{\frac{\tau(1{-}\varepsilon ) {+}(\varepsilon {+}1) \left[1{+}(\tau {-}1) \chi ^2\right] }{\tau -1{+}\varepsilon  (\tau {-}1) \left(\chi ^2{-}1\right){+}(\tau {+}1) \chi ^2}} ,
\end{equation}
and the second one $\varepsilon_{e} \geq 1$, which corresponds to
\begin{equation} 
1 \leq g \leq g_{\varepsilon}= \sqrt{\frac{\left(1{-}\varepsilon ^2\right) (1{-}\tau ){+}2 \sqrt{(\varepsilon ^2{-}1) \tau }}{(\varepsilon {-}1) \left[\tau {+}1{+}\varepsilon  (\tau {-}1)\right]}} ,
\end{equation}
and so the maximum attainable gain is
\begin{equation} 
g_{\text{max}}=\min \{ g_{\chi}, g_{\varepsilon}  \} \,.
\end{equation}

Thus, the overall condition for error correction is given by
\begin{equation} 
g_{\text{max}} > 1/\chi \,.
\label{errcondition}
\end{equation}

In Fig.~\ref{figapp2} of the Appendix \ref{ap2}, we provide the set of channels that can be error corrected with this protocol for different resource states, taking into account both the maximum attainable NLA gain, $g_{\text{max}} > 1/\chi $, and the entanglement-breaking condition, $v \geq 1 + |\tau|$. As expected, the success of the error correction protocol is intimately related with the entanglement power of the resource state.

The consistent behavior observed before, highlighted by the relationship between the entanglement of the resource and the output state, where the error correction begins when the resource state reaches the deterministic upper bound of entanglement, is based on the measure used in this analysis, i.e., entanglement of formation. If, instead, we had picked logarithmic negativity as the entanglement quantifier, then there is no apparent connection between the entanglement of the resource state and the error correction protocol. In other words, using logarithmic negativity we can easily find situations where the deterministic upper bound of entanglement has been surpassed while the overall output has not been error corrected. That implies that from this point of view logarithmic negativity in general overestimates the entanglement which might lead to erroneous conclusions when it is used for example in distillation protocols giving a potentially false positive result. Reaching that physical bound with entanglement of formation is a significantly harder task, but as soon as we have reached it then the distilled entangled state is objectively more entangled since it is useful to perform tasks such as error correction discussed above.

\subsection{Error Correction with realistic NLA}

When the NLA is experimentally implemented with linear optics, it consists of $N$ modified quantum scissors devices \cite{Pegg.Phillips.Barnett.PRL.98} (schematically presented in Appendix \ref{ap2}, Fig.~\ref{figapp3}). The input state is split on an array of beam-splitters with each mode then being passed through an individual quantum scissor. The modes are then coherently recombined to form the output state, with the correctly amplified state being achieved only when each quantum scissor heralds successful operation.

In the ideal case, the NLA operation is given by $g^{\hat{n}} $. However, using $N$ quantum scissors the corresponding operation is given by \cite{Ralph.Lund.QCMC.09,Dias.Ralph.PRA.17}
\begin{equation} 
\hat{T}_N := \hat{\Pi}_N  g^{\hat{n}} \,,
\end{equation}
where $\hat{\Pi}_N$ is the truncation operation defined as
\begin{equation} 
\hat{\Pi}_N := \left(\frac{1}{1+g^2} \right)^{N/2}\sum_{n=0}^N \frac{N!}{(N-n)!N^n} \ket{n}\bra{n} \,.
\end{equation}

This operation leads to a state truncation in the photon number basis to order $N$, with $g = \sqrt{\left(1 - \xi \right) /\xi}$ being a gain controlled by a tunable beam-splitter ratio, $\xi$. The physical construction of the NLA reduces to the ideal one only in the unphysical asymptotic limit of $N \rightarrow \infty $.

Successful operation of the NLA decreases exponentially with $N$, so we consider the case where the NLA consists of the simplest setup, i.e., a single quantum scissor. The reason for this is two-fold, it represents the simplest experimental setup and maximizes probability of success. The single quantum scissor induces the following transformation 
\begin{equation}
\hat{T}_1(\alpha\ket{0} + \beta\ket{1} + \gamma  \ket{2} + \cdots) =\frac{\alpha\ket{0} + g\beta \ket{1}}{ \sqrt{1+g^2}} \,,
\end{equation}
with all higher order terms being truncated. The effect of this truncation is to introduce a small amount of excess noise to the output state. Due to the nature of this operation, the effect of this truncation on large amplitude input states is severe and will result in a large amount of excess noise. Therefore, when the NLA is implemented with a single quantum scissor, the error correction protocol performs best in the high-loss regime.

\begin{figure}[t]
\centering
  \includegraphics[width=\columnwidth]{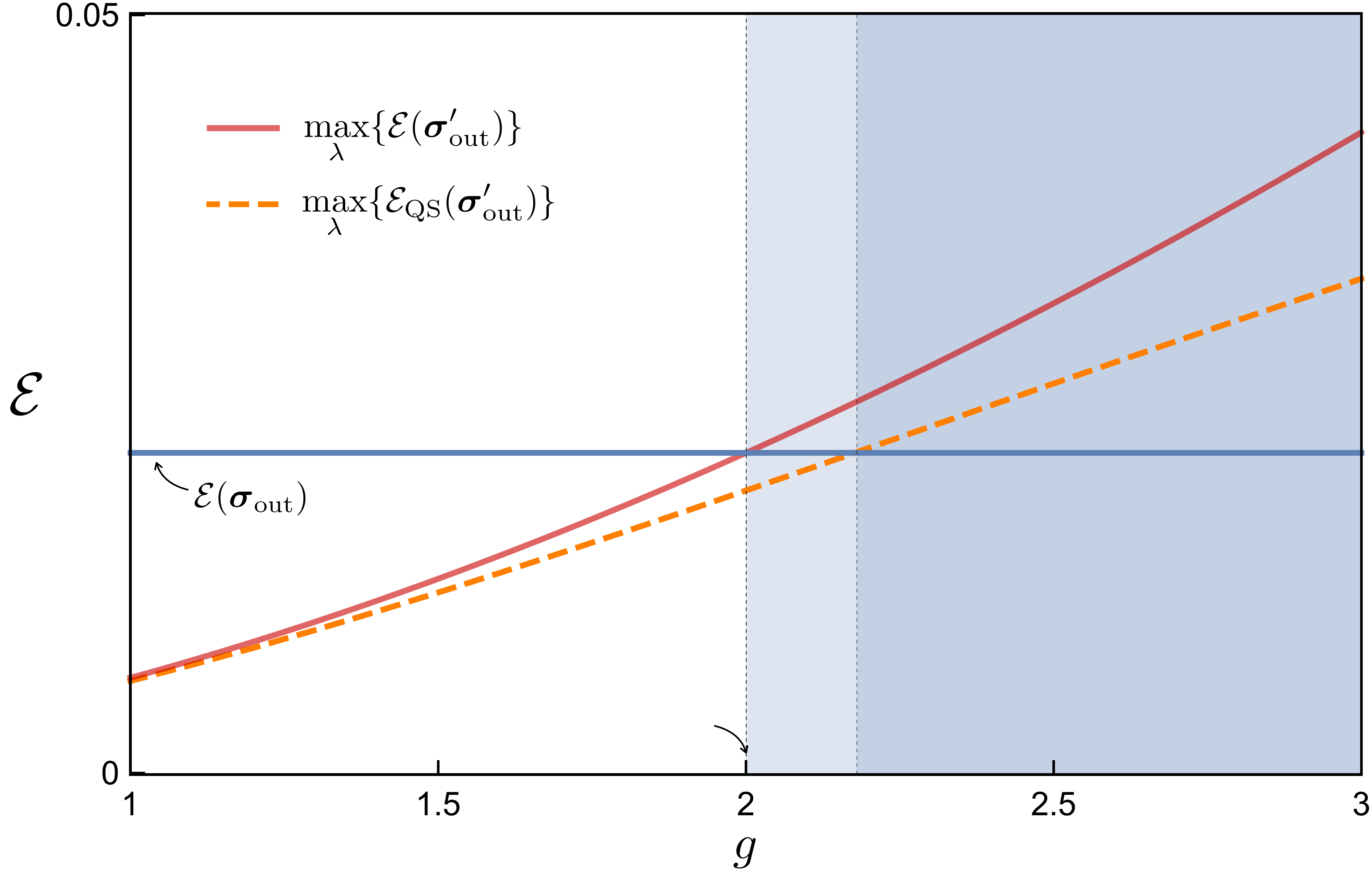}
  \caption{ \small Error correction with a quantum scissor. The thermal loss channel that we want to error correct has transmissivity $\tau=0.01$ and noise $\varepsilon=1.0002$. Both the resource and the initial state have squeezing parameter equal to $\chi=\zeta=0.5$. The red solid line depicts $\max_{\lambda}\{\mathcal{E}(\boldsymbol{\sigma}'_{\text{out}})\}$ for the ideal NLA, and the brown dashed one depicts the corresponding $\max_{\lambda}\{\mathcal{E}_{\text{QS}}(\boldsymbol{\sigma}'_{\text{out}})\}$ for the realistic NLA with one quantum scissor. As we see the NLA gain needed to cross the value $\mathcal{E}(\boldsymbol{\sigma}_{\text{out}})$ is greater for the realistic NLA compared with the ideal one. All quantities plotted are dimensionless.}
  \label{fig5}
\end{figure}

In Fig.~\ref{fig5} we show the difference between the NLA gain needed for error correction for both the realistic and the ideal NLA implementations. As it is expected the real NLA needs a higher gain compared to the ideal one.

\section{Conclusions}

In conclusion, we showed that every phase-insensitive Gaussian channel can be simulated through teleportation with a physical resource state (Eqs.~\ref{physicalstates1}\,-\ref{physicalstates3}), and we derived analytical expressions for all the possible states able to perform this task. We also identified the optimal resource states (Eq.~\ref{optstates}), i.e., the states with the minimum requirements in energy and entanglement, using entanglement of formation as the entanglement measure. This result clarifies a previously published work \cite{Scorpo.et.al.PRL.17}, where, using logarithmic negativity as the entanglement quantifier, pure channels were not able to be simulated with the optimal resource states. This discrepancy is not surprising, since logarithmic negativity is not a proper entanglement measure (even though it is an entanglement monotone \cite{Plenio.PRL.05}), and it's not the first time that its problematic (inconsistent) behavior has been observed \cite{Tserkis.Ralph.PRA.17,Wolf.Giedke.Cirac.PRL.06}. We also showed that resource states with entanglement equal to the Choi-state can simulate channels that decohere an incoming state in the same way as the initial one, and we used that fact to generalize an error correction protocol for noise induced by thermal loss channels.

The next step is to extend this error correction protocol to other useful Gaussian channels. We should also note that recently this finite-energy analysis of resource states has found practical applications to private communication \cite{Kaur.Wilde.PRA.17,Laurenza.Braunstein.Pirandola.PRA.17,Scorpo.Adesso.SPIE.17}. Finally, the consistent behavior of entanglement of formation identified here (see Eq.~\ref{optstates} and Fig.~\ref{fig4}) implies that it can also be used as a faithful quantifier in concatenated error correction protocols, such as quantum repeaters \cite{Sangouard.et.al.RVP.11,Furrer.Munro.arXiv.16,Dias.Ralph.PRA.17}, where the incoming state of the first part becomes a resource state of the next part and so on.

\section*{Acknowledgements}

We thank Nedasadat Hosseinidehaj for proofreading the first version of the paper. The research is supported by the Australian Research Council (ARC) under the Centre of Excellence for Quantum Computation and Communication Technology (CE110001027).

\appendix
\renewcommand{\thefigure}{A\arabic{figure}}

\section{Problems with Fidelity}
\label{ap1}
\setcounter{figure}{0}
\begin{figure}[b] 
\centering
  \includegraphics[width=\columnwidth]{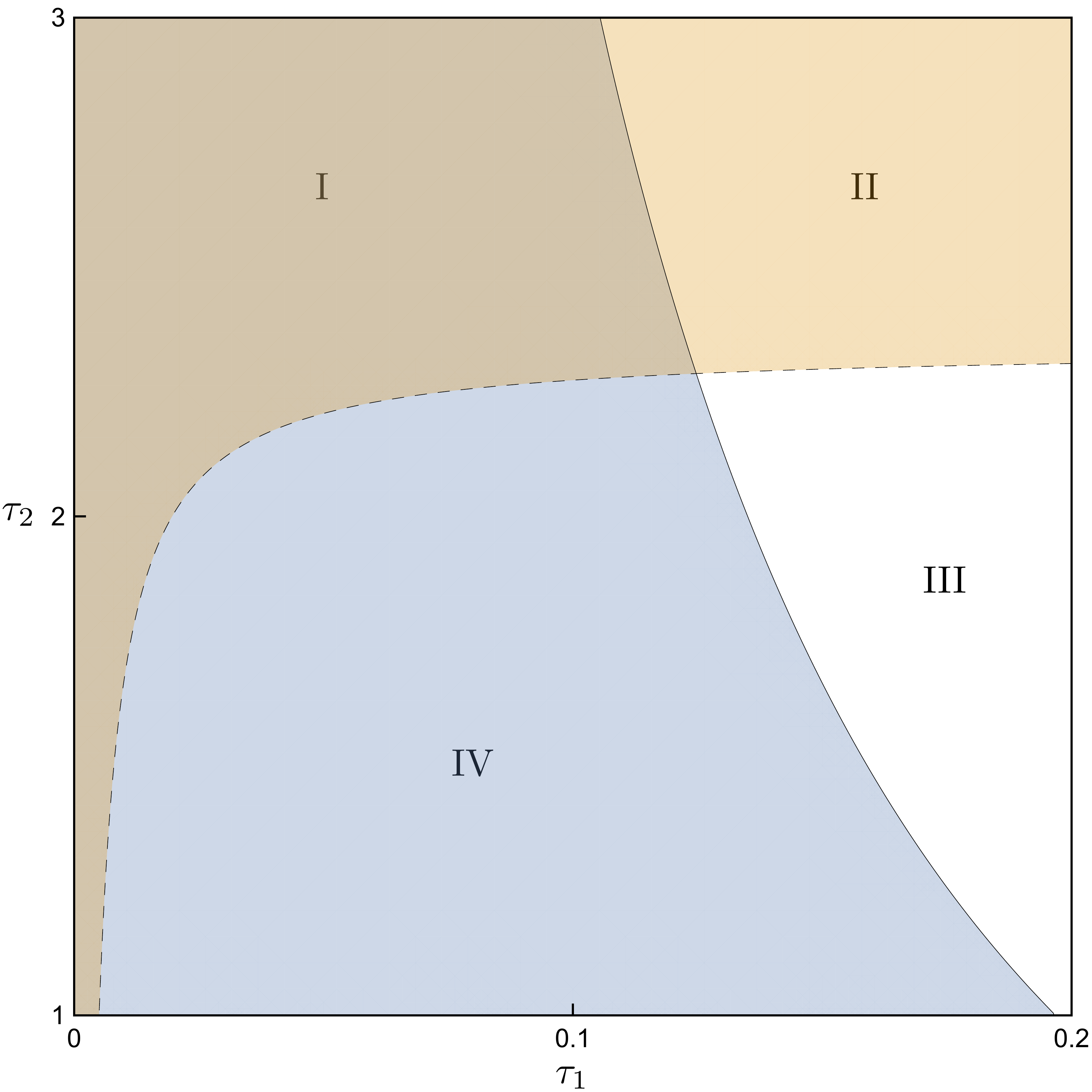} 
  \caption{ \small Problems with fidelity. Let us assume that a pure state with squeezing parameter equal to $\zeta=0.8$ is going through both a thermal amplifier channel $\mathcal{A}(\tau_2,\varepsilon_2=2.5)$ and a thermal loss channel $\mathcal{L}(\tau_1,\varepsilon_1=1.01)$. For different values of $\tau_1$ and $\tau_2$ we calculate the fidelity of the input/output state and we get two sets of states: (i) the set with $\mathcal{F}_1<\mathcal{F}_2$, colored with blue (on the left of the solid line, sections I and IV) and (ii) the ones with no entanglement left, colored with brown (on the top of the dashed line, sections I and II). As we can see there is an overlap between those two sets, i.e., section I, where we have both an entanglement-breaking situation and $\mathcal{F}_1<\mathcal{F}_2$. All quantities plotted are dimensionless.}
  \label{figapp1}
\end{figure}

Assuming that we have two quantum states with density matrices $\boldsymbol{\rho}$ and $\boldsymbol{\sigma}$, one way to measure how close one state is to the other is fidelity \cite{Uhlmann.RMP.76,Marian.Marian.PRA.12,Barnum.et.al.PRL.96,Banchi.Braunstein.Pirandola.PRL.15}, defined as
\begin{equation} 
\mathcal{F}(\boldsymbol{\rho},\boldsymbol{\sigma}):=\left(\tr\sqrt{\sqrt{\boldsymbol{\rho}}\,\boldsymbol{\sigma}\sqrt{\boldsymbol{\rho}}}  \right)^2 .
\end{equation}

Fidelity has extensively been used in quantum information as an indicator of a successful protocol. Even though this measure gives a physical insight, it should be used with caution, since comparing non-unity fidelities does not necessarily provide enough evidence that a state is ``closer" to the target one, if by closer we want to take into account properties like classicality, separability, Gaussianity etc. More specifically, if a target state is an entangled one, and this entanglement is crucial to the protocol, then finding a separable state which has fidelity with the target state close to unity doesn't help at all. 

Let us for instance assume that we send a state through an entanglement-preserving thermal loss channel $\mathcal{L}(\tau_1,\varepsilon_1)$ and we calculate the fidelity between the input and output state, $\mathcal{F}_1$. Now, before the thermal loss channel, we introduce an amplifier channel, $\mathcal{A}(\tau_2,\varepsilon_2)$, in way that the whole channel $\mathcal{A}(\tau_2,\varepsilon_2) {\circ} \mathcal{L}(\tau_1,\varepsilon_1)$ is entanglement-breaking, and we calculate again the fidelity, $\mathcal{F}_2$. We might expect that always $\mathcal{F}_1>\mathcal{F}_2$, since for the second case all the quantum correlations are gone and the state passing through an entanglement-breaking channel is useless for quantum communication, but we can easily find certain channels $\mathcal{L}$ and $\mathcal{A}$ for which $\mathcal{F}_1<\mathcal{F}_2$. A specific example is presented in Fig.~\ref{figapp1}. 

\section{Noiseless Linear Amplification}
\label{ap2}
\subsection{Effective parameters of the NLA}

The effective parameters for a thermal loss channel $\{ \chi_e,\tau_e,\varepsilon_e\}$ have been computed in Refs.~\cite{Walk.Lund.Ralph.13,Blandino.et.al.PRA.15}, and are given by
\begin{subequations}
\begin{align} 
\chi_{e}&=\chi  \sqrt{\frac{2{+}(g^2{-}1) \left[\varepsilon  (\tau {-}1){+}\tau {+}1\right]}{2{+}(\varepsilon{-}1) (\tau {-}1) (g^2{-}1)}}  , \\
\tau_{e}&=\frac{4 g^2 \tau }{\varepsilon{+}1 {+}(\varepsilon {-}1) \left[ (\tau {-}1)g^2{-}\tau \right]}  \nonumber \\
&\hspace{0.5cm}\times \frac{1}{(\varepsilon {+}1) (1{-}\tau ){+}\left[\tau {+}1{+}\varepsilon  (\tau {-}1)\right]g^2 } ,\\
\varepsilon_{e}&=\frac{\tau {+}1{+}\varepsilon  \left[2{+}\varepsilon  (1{-}\tau )\right]{+}(\varepsilon {-}1)  \left[\tau {+}1{+}\varepsilon  (\tau {-}1)\right]g^4}{\left[\varepsilon{+}1 {-}(\varepsilon {-}1) g^2\right]^2{-}\tau\left(\varepsilon ^2{-}1\right) \left(g^2{-}1\right)^2  }  .
\end{align}
\end{subequations}

\subsection{Maximum gain of the NLA}

\begin{figure}[t]
\centering
  \includegraphics[width=\columnwidth]{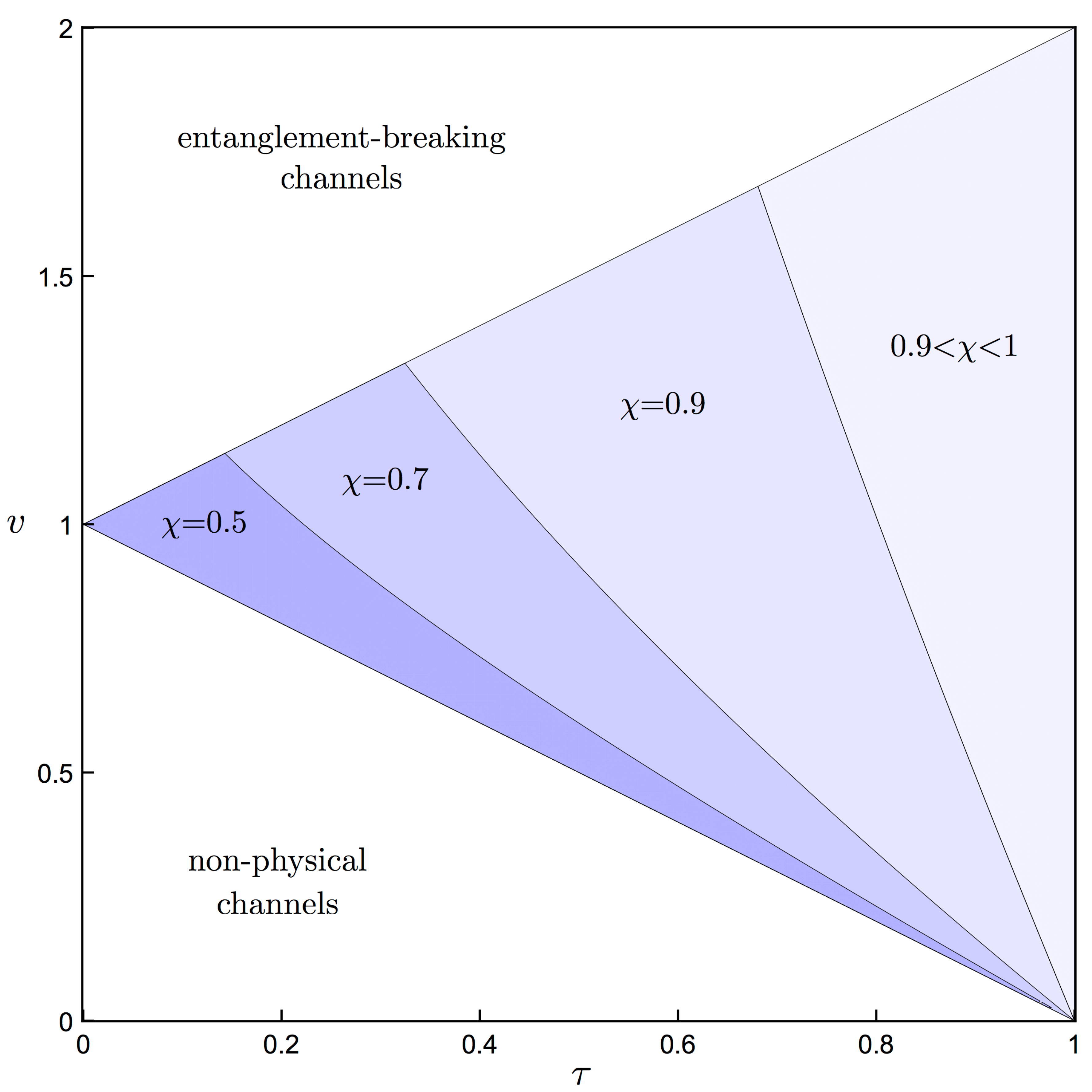}
  \caption{ \small Error correction range. We plot the range of all the possible channels with parameters $0 \leq \tau \leq 1$ and $1\leq v\leq 2$ that can be error corrected with the protocol, based on both the NLA condition, $g_{\text{max}} > 1/\chi $, and the entanglement-breaking condition, $v\geq 1 + |\tau|$. It is apparent that for increasing values of squeezing parameter $\chi$, the set of channels is increased as well. All quantities plotted are dimensionless.}
  \label{figapp2}
\end{figure}

In Fig.~\ref{figapp2}, we present the range of channels that can be error corrected for resource states with different squeezing parameters, $\chi$. We take into account both the maximum attainable NLA gain, $g_{\text{max}} \geq 1/\chi $, and the entanglement-breaking condition, $v \geq 1 + |\tau|$. 

\subsection{NLA with quantum scissors}

\begin{figure}[t]
\centering
  \includegraphics[width=\columnwidth]{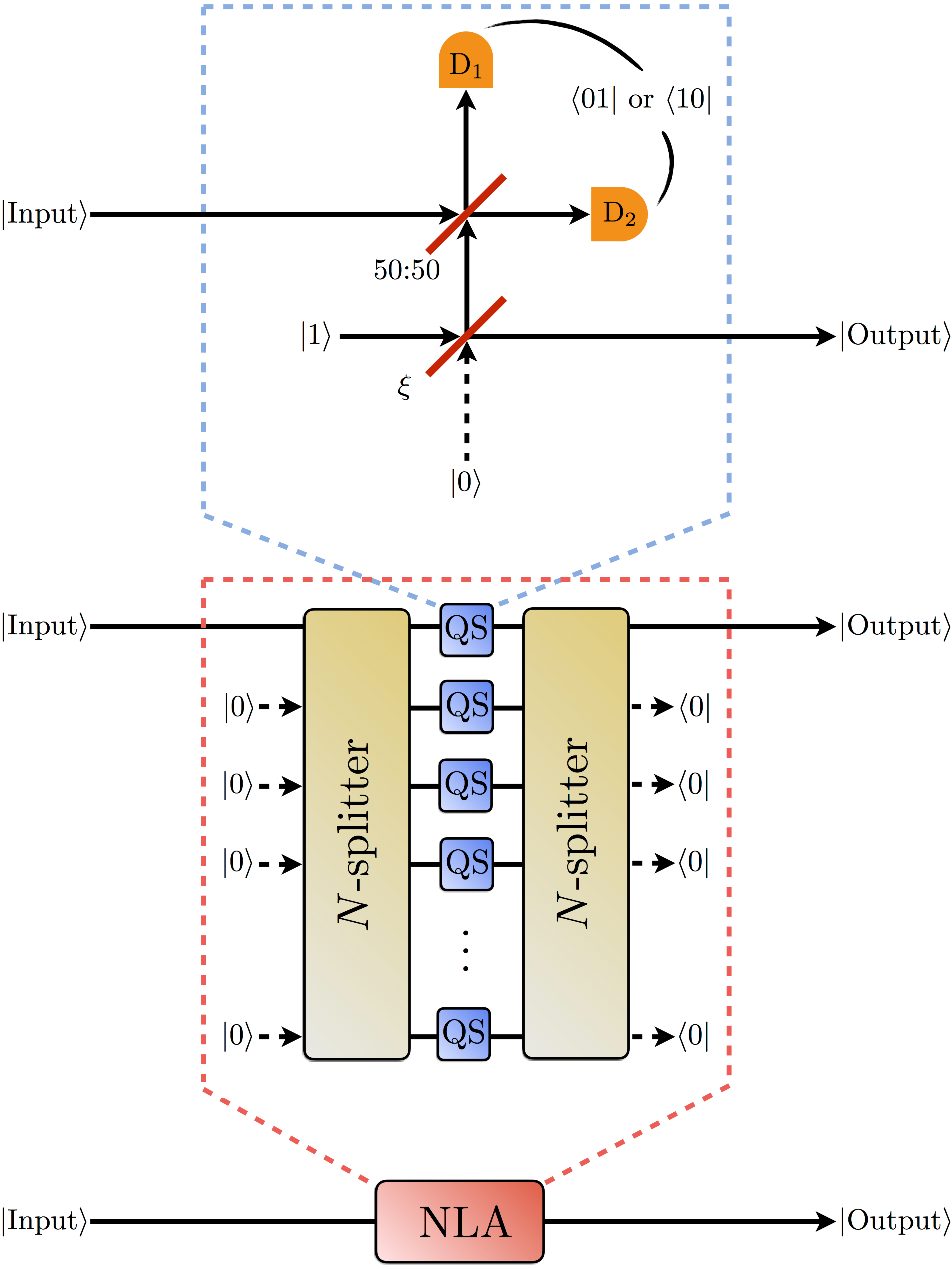}
  \caption{ \small NLA with quantum scissors. Each quantum scissor operation, QS, consists of two beam-splitters. The input signal is mixed on a balanced beam-splitter with an ancilla signal and both outputs are measured using photon detectors, $\text{D}_1$ and $\text{D}_2$. The ancilla signal is one of the two outputs of a single photon passing through a tunable beam-splitter with ratio $\xi$, while the other signal is the overall output of the quantum scissor. Successful quantum scissor operation is heralded when a single photon is detected at $\text{D}_1$ and none at $\text{D}_2$ or vice versa. Using $N$ quantum scissors and two $N$-splitters (one to divide and another to recombine the signal) we can approximate the ideal NLA in the limit of $N \rightarrow \infty $.}
  \label{figapp3}
\end{figure}

In Fig.~\ref{figapp3} we present the components of a quantum scissor, and how noiseless linear amplification can be constructed using an array of $N$ quantum scissors.

\end{document}